\documentclass[letter]{aa}

\usepackage{graphicx}
\usepackage{hyperref}
\usepackage{natbib}

\newcommand{\cross}[1][1pt]{\ooalign{%
  \rule[0.7ex]{0.7ex}{0.15ex}\cr
  \hss\rule{0.15ex}{.5em}\hss\cr}}

\usepackage{marvosym}
\bibpunct{(}{)}{;}{a}{}{,}
\usepackage{txfonts}
\def\micron{$\mu$m}
\def\deg{$^\circ$}
\def\arcsec{"}

\def\aap{A\&A}

\def\apjl{ApJL}

\def\kms{${\rm km}\,{\rm s}^{-1}$}
\def\lesssim{\mathrel{\hbox{\rlap{\hbox{\lower4pt\hbox{$\sim$}}}\hbox{$<$}}}}
\def\gtrsim{\mathrel{\hbox{\rlap{\hbox{\lower4pt\hbox{$\sim$}}}\hbox{$>$}}}}

\def\rsun{R$_{\odot}$\,}
\def\msun{M$_{\odot}$\,}

\begin{document}
  
\title{V838\,Monocerotis: the central star and its environment a
  decade after outburst}

\author{O. Chesneau\inst{1,\cross} 
  \and
  F. Millour\inst{1}
  \and
  O. De Marco\inst{2}
  \and
  S. N. Bright\inst{1,2}
  \and
  A.~Spang\inst{1}		
  \and
  D.~P.~K. Banerjee\inst{3}
  \and
  N.~M.~Ashok\inst{3}
  \and
  T.~Kami\'nski\inst{4}
  \and
  J~P. Wisniewski, \inst{5}
  \and
  A. Meilland\inst{1}
  \and
  E. Lagadec\inst{1}
  \thanks{Based on observations at Paranal
    Observatory under programs 088.D-0005, 090.D-0011, 091.D-0030 and
    093.D-0056.\protect \\ $^{\cross}$ O. Chesneau passed away
    shortly before submitting this letter. We express our profound
    sadness on this untimely demise and convey our deepest condolences
    to his family.}  
}

\offprints{Florentin.Millour@oca.eu}

\institute{ Laboratoire Lagrange, UMR7293, Univ. Nice
  Sophia-Antipolis, CNRS, Observatoire de la C\^ote d'Azur, 06300
  Nice, France
  \and
  Department of Physics \& Astronomy, Macquarie University, Sydney,
  NSW 2109, Australia
  \and
  Physical Research Laboratory, Navrangpura, Ahmedabad, Gujarat, India
  \and 
  Max-Planck Institut f\"{u}r Radioastronomie, Auf dem H\"{u}gel 69,
  D-53121 Bonn, Germany
  \and
  HL Dodge Department of Physics \& Astronomy, University of Oklahoma,
  440 W Brooks Street, Norman, OK 73019, USA
}

\date{Received, accepted.}

\abstract {}
{V838 Monocerotis erupted in 2002, brightened in a series of
  outbursts, and eventually developed a spectacular light echo. A very
  red star emerged a few months after the outburst. The whole event
  has been interpreted as the result of a merger.}
{We obtained near-IR and mid-IR interferometric observations of
  \object{V838\,Mon} with the {{\sc AMBER}} and {{\sc MIDI}}
  recombiners located at the Very Large Telescope Interferometer
  ({{\sc VLTI}}) array. The {{\sc MIDI}} two-beam observations were
  obtained with the 8m Unit Telescopes between October 2011 and
  February 2012. The {{\sc AMBER}} three-beam observations were
  obtained  with the compact array (B$\leq$35m) in April
  2013 and the long array (B$\leq$140m) in May 2014, using the 1.8m
  Auxiliary Telescopes.}
{A significant new result is the detection of a compact structure
  around \object{V838\,Mon}, as seen from {{\sc MIDI}} data. The
  extension of the structure increases from a FWHM of 25\,mas at
  8\,\micron\ to 70\,mas at 13\,\micron.
  At the adopted distance of $D = 6.1 \pm 0.6$\,kpc, the dust is
  distributed from about 150 to 400\,AU around \object{V838\,Mon}. The
  {{\sc MIDI}} visibilities reveal a flattened structure whose aspect
  ratio increases with wavelength. The major axis is roughly oriented
  around a position angle of $-10$\deg, which aligns with previous
  polarimetric studies reported in the
  literature. This flattening can be interpreted as a relic of the
  2002 eruption or by the influence of the currently embedded B3V
  companion. The {{\sc AMBER}} data provide a new diameter for the
  pseudo-photosphere, which shows that its diameter has decreased by
  about 40\% in 10yrs, reaching a radius $R_{*} = 750 \pm
  200$\,\rsun\ ($3.5 \pm 1.0$\,AU).
}
{After the  2002 eruption, interpreted as the merging of two 
  stars, it seems that the resulting source is relaxing to a normal
  state. The nearby environment exhibits an equatorial over-density of
  dust up to several hundreds of AU. }

\titlerunning{V838\,Mon 10 years after outburst}
\authorrunning{Chesneau et al.}

\maketitle

\section{Introduction}

\object{V838\,Mon} is a nova-like object, which erupted in 2002 in a
series of outbursts (reaching V$\sim$6.8). It subsequently developed a
light echo which was extensively studied by
\citet{2003Natur.422..405B, 2006ApJ...644L..57B,
  2012A&A...548A..23T}. The eruption was unlike classical novae as the
effective temperature of the object dropped and the spectral type
evolved into a very late L-type supergiant \citep{2003MNRAS.343.1054E,
  Loebman2014}. The event has been interpreted as the merger of a
$\sim$8\,\msun star with a sub-solar mass star
\citep{2003ApJ...582L.105S,2005A&A...441.1099T,
  2006A&A...451..223T}. The mid infrared flux of V838 Mon increased by
a factor of two between 2004 and 2007, suggesting that new dust was
forming in the expanding ejecta of the outbursts
\citep{2003ApJ...598L..43W, 2005A&A...436.1009T,
  2008ApJ...683L.171W}. The expanding ejecta engulfed a 
companion close to the central source \citep{2006ATel..966....1B,
  2007A&A...474..585M, 2009ATel.2211....1K}.

The progenitor of \object{V838\,Mon} was found in a variety of
surveys, including 2MASS, and appears consistent with a somewhat
reddened early main sequence star. Post-outburst observations have
found a faint, blue component in the spectrum and
\citet{2005A&A...441.1099T} argued that the pre-outburst spectral
energy distribution (SED) is well-matched by a pair of early main
sequence stars (B3V + B1.5V or B4V + A0.5V), making V838 Mon a triple
system {\it de facto} (main star + sub-solar merging star + B3V
companion). \citet{2004ApJ...607..460L} obtained numerous IR spectra
of \object{V838\,Mon} during and after the eruption (2002-2003). They
fit their data to a model consisting of a cool (T$_{\rm eff}$ = 2100K,
R$_* = 8.8$\,AU) stellar photosphere surrounded by a large, absorbing
molecular cloud. A more recent work \citep{Loebman2014} provides
precise estimates of the stellar and enshrouding cloud parameters,
with an effective temperature of the star of 2000--2200\,K, and a
radius of the shell of $R=263\pm10$\,AU.

\citet{2005ApJ...622L.137L} first used long-baseline near-IR
(2.2\,\micron\, baselines smaller than 85\,m) interferometry in
November-December 2004 to provide the first direct measurement of the
angular size of \object{V838\,Mon} using the two-telescope recombiner
Palomar Testbed Instrument ({{\sc PTI}}). At this epoch, they measured 
an angular diameter for the central source of $\Theta = 1.83 \pm
0.06$\,mas.
Reasonably accurate distances were determined from the light echo and
using other methods \citep{2005A&A...434.1107M, 2008AJ....135..605S,
  2011A&A...529A..48K}. We adopt the distance $D = 6.1 \pm 0.6$\,kpc
from \citet{2008AJ....135..605S}. Using this distance, the
interferometric diameter translates to a linear radius for the
supergiant of $1200 \pm 150$\,\rsun\ ($5.6 \pm
0.7$\,AU). \citet{2005ApJ...622L.137L} could also fit the data
assuming an elliptical structure with a major axis oriented at
P.A. $15$\deg\ whose extensions are $3.57 \times 0.07$\,mas (i.e. the
minor axis is unresolved).

\object{V838\,Mon} is embedded in a dense, large scale environment
\citep{2012A&A...548A..23T, 2011A&A...529A..48K, 2014apn6.confE.118E},
potentially contaminating observations of the central star. High
angular resolution studies with a field-of-view (FOV) of a few arcsec
are an asset in that domain to isolate the central regions from the
extended dusty cloud. A great advantage of optical interferometry
studies is to fully isolate the measurements from the extended
environment.

This letter presents optical interferometry measurements obtained with
the Very Large Telescope Interferometer (VLTI). The observations are
presented in Sect. \ref{Observations}. In Sect. \ref{Analysis} we
analyze the {{\sc MIDI}} mid-IR measurements followed by the {{\sc
    AMBER}} $K$ band measurements by means of simple geometrical
models. The results are then discussed in Sect. \ref{Discussion}.


\section{Observations}
\label{Observations}

Mid-IR interferometric data were obtained with the two-telescope
recombiner {{\sc MIDI}} \citep{2004A&A...423..537L} between October
2011 and February 2012 using the 8m Unit Telescopes (UTs) of the Very
Large Telescope. These observations provided dispersed fringe
visibilities between 8 and 13\,\micron\ in addition to classical
spectrophotometric capabilities.

Near-infrared observations were obtained with {\sc AMBER} \citep[a
  three-telescopes combiner located at the
  VLTI:][]{2007A&A...464....1P} in November 2012 and February 2013,
using the low spectral resolution mode (R=35) with the
UTs. Unfortunately, these data are unusable due to a bad calibrator
(see details in appendix \ref{sec:badcal}) and therefore not presented
here.
In April 2013, observations were repeated with the 1.8m Auxiliary
Telescopes (ATs) under photometric conditions, using the compact
configuration ($B\leq35$m) and more suitable calibrators. The extended
configuration ($B\leq140$m) was subsequently used in May 2014 to
observe \object{V838 Mon}.

\begin{figure}[htbp]
  \centering
  \includegraphics[width=0.22\textwidth]{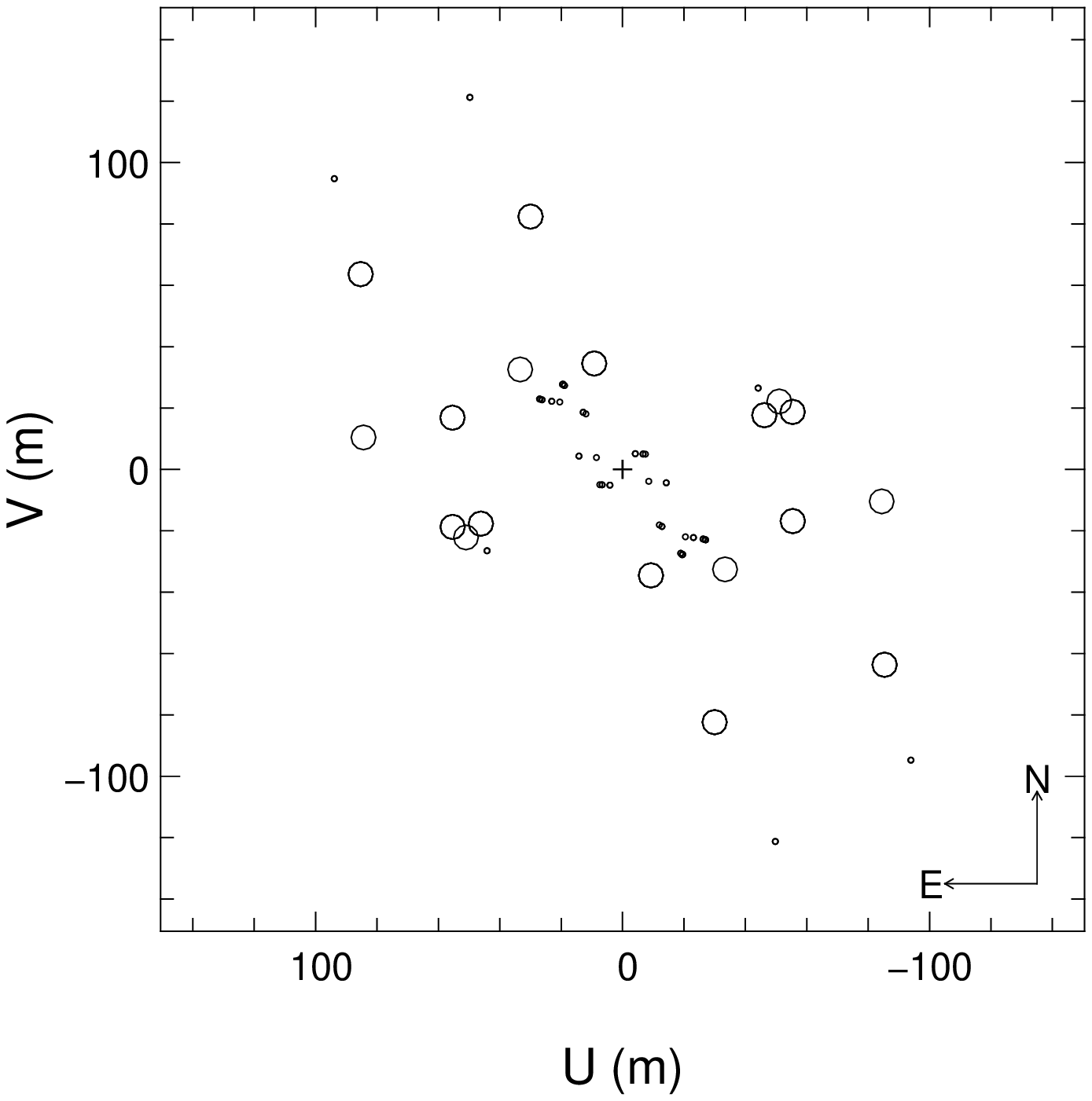}
    \hspace{5mm}
    \includegraphics[width=0.22\textwidth]{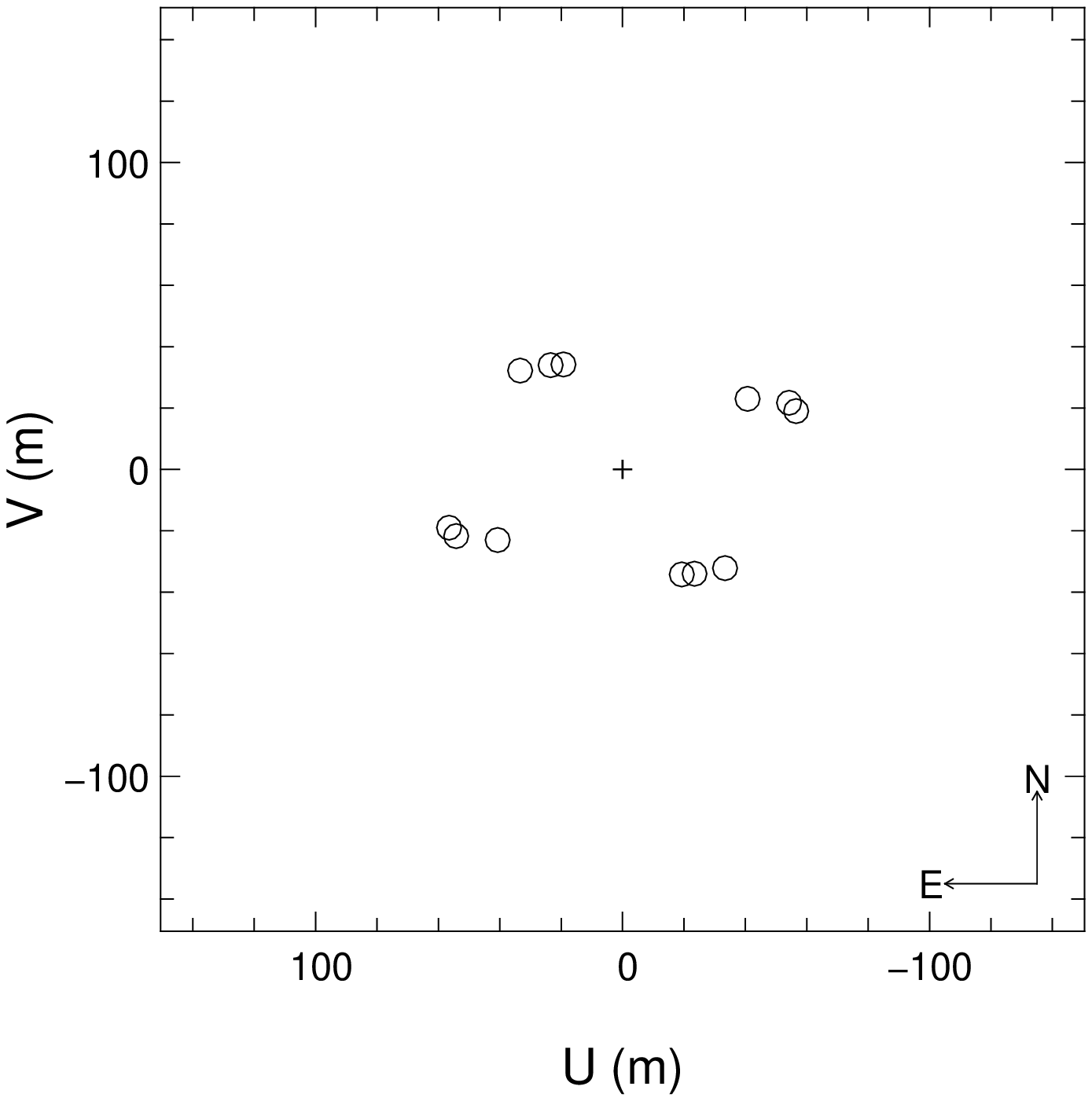}
  \caption{{\bf Left:} ($u,v$) coverage of the AMBER observations. Small
    dots are observations with ATs and larger circles with UTs.
    {\bf Right:} ($u,v$) coverage of the MIDI observations.
    \label{fig:AMBER_uv}}
\end{figure}

The great advantage of both instruments is that the FOV of the
interferometric measurements are within the point spread function of a
single telescope, i.e. the same value of $\sim$0.3\arcsec\ at
2.2\,\micron\ for the 1.8m ATs with {{\sc AMBER}}, and at
10\,\micron\ for the 8m UTs with {\sc MIDI}. This isolates the central
source from the extended emission of the cloud in which
\object{V838\,Mon} is embedded. The observation log is presented in
Table\,\ref{tab:logObs} and the ($u,v$) plane coverage is plotted in
Fig.\ref{fig:AMBER_uv}.

We reduced the {{\sc AMBER}} data using the standard data reduction
software {\tt amdlib v3.0.8} \citep{2007A&A...464...29T,
  2009A&A...502..705C}. Near-infrared $JHK$ photometry was obtained on
a regular basis from the 1.2m telescope at the Mt. Abu Observatory,
India \citep{Banerjee2012ca}. These measurements helped to prepare the
interferometric observations. On March 1, 2013, the $J$, $H$, $K$ band
magnitudes were 7.12, 5.91, 5.43, respectively, almost unchanged since
November 2011.

\begin{table}[htbp]
  \caption[]{Log of {{\sc AMBER}} and {{\sc MIDI}} observations of
    \object{V838\,Mon}.}
  \label{tab:logObs}
  \centering
  \begin{tabular}{lcccccccc}
    \hline
    \noalign{\smallskip}
    Date & Stations & Calibrators & Wavelength & Nb. obs. \\ 
    \noalign{\smallskip}
    \hline
    \noalign{\smallskip}
    {\bf MIDI}\\
    10/10/2011 &  U3-U4 & 1$^\circ$  & 8--13\,\micron &  2 \\
    13/12/2011 &  U2-U3 & 1$^\circ$  & 8--13\,\micron &  1 \\
    14/12/2011 &  U2-U3 & 1$^\circ$  & 8--13\,\micron &  1 \\
    11/02/2012 &  U3-U4 & 1$^\circ$  & 8--13\,\micron &  1 \\
    {\bf AMBER}\\
    14/04/2013 & A1-B2-D0 & 6$^{\lhd}$, 7$^{\rhd}$ & 1.54--2.50\,\micron &  3 \\
    15/04/2013 & A1-C1-D0 & 6$^{\lhd}$, 8$^\bigtriangleup$ & 1.56--2.50\,\micron &  2 \\
    06/05/2014 & A1-G1-J3 & 9$^\bigtriangledown$        & 1.53--2.46\,\micron &  1 \\
    \noalign{\smallskip}
    \hline
  \end{tabular}
  \tablefoot{\tiny Calibrator angular diameters from SearchCal/JMMC
    \citep{2006A&A...456..789B} and getCal/NExScI.
    $^\circ$: HD\,52666 $2.53\pm 0.21 $\,mas,
    $^\lhd$: HD\,52265 $0.44 \pm 0.03$\,mas, 
    $^\rhd$: HD\,64616 $0.51 \pm 0.04$\,mas, 
    $^\bigtriangleup$: HD\,63660 $0.96 \pm 0.07$\,mas, 
    $^\bigtriangledown$: HD\,54990 $0.57 \pm 0.59$\,mas.}
\end{table}

\begin{figure}[htbp]
  \centering
  \includegraphics[height=0.4\textwidth,angle=-90]{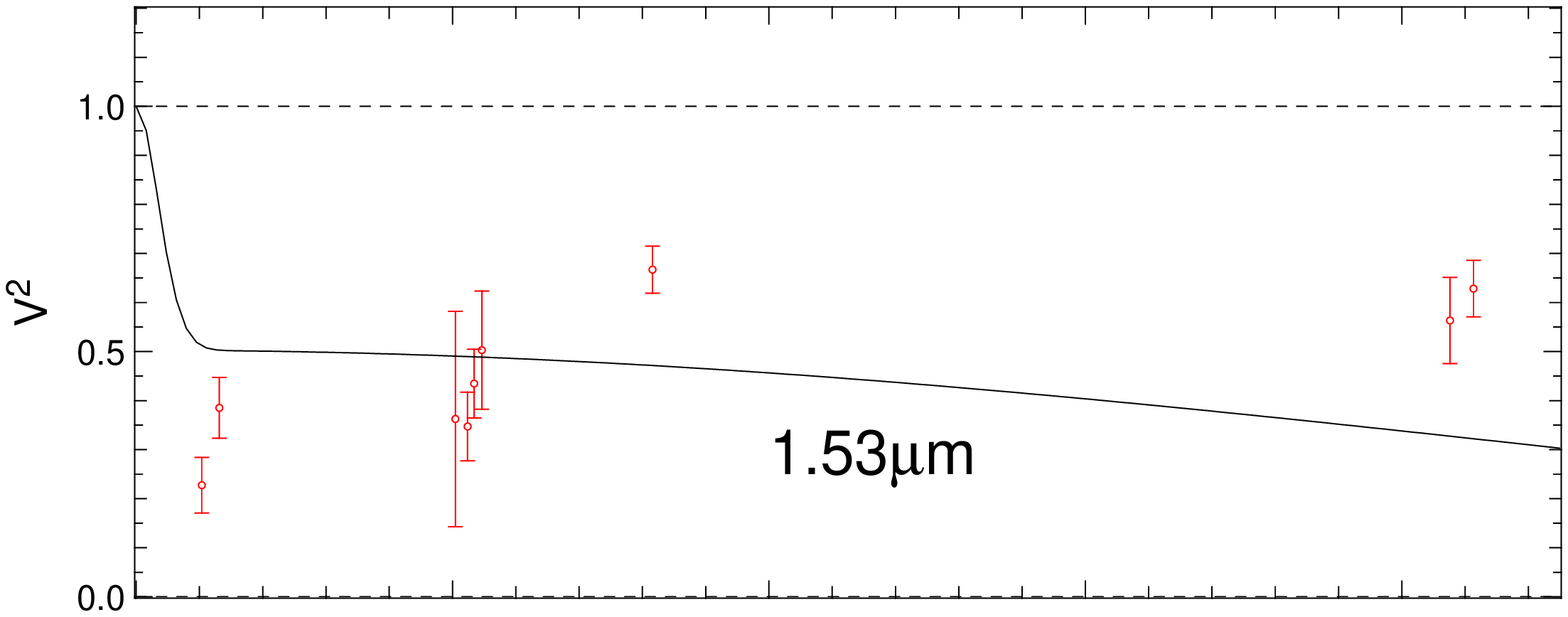}
  \includegraphics[height=0.4\textwidth,angle=-90]{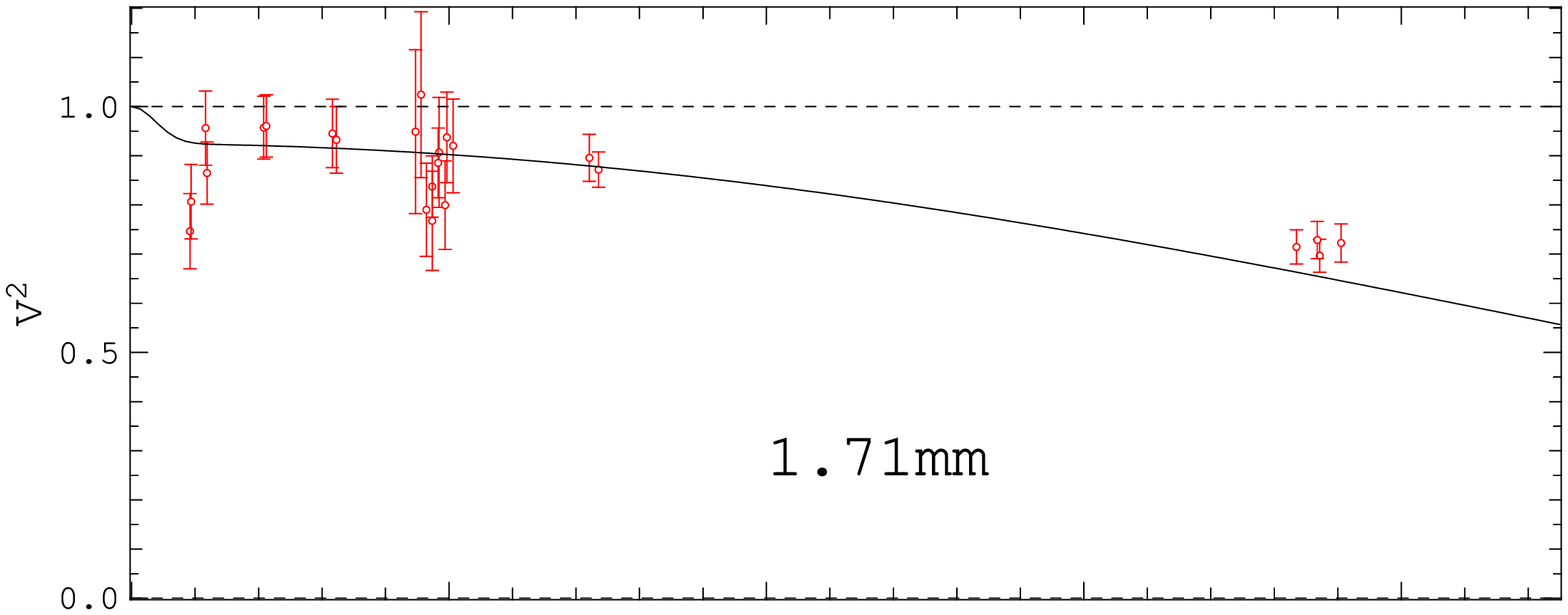}
  \includegraphics[height=0.4\textwidth,angle=-90]{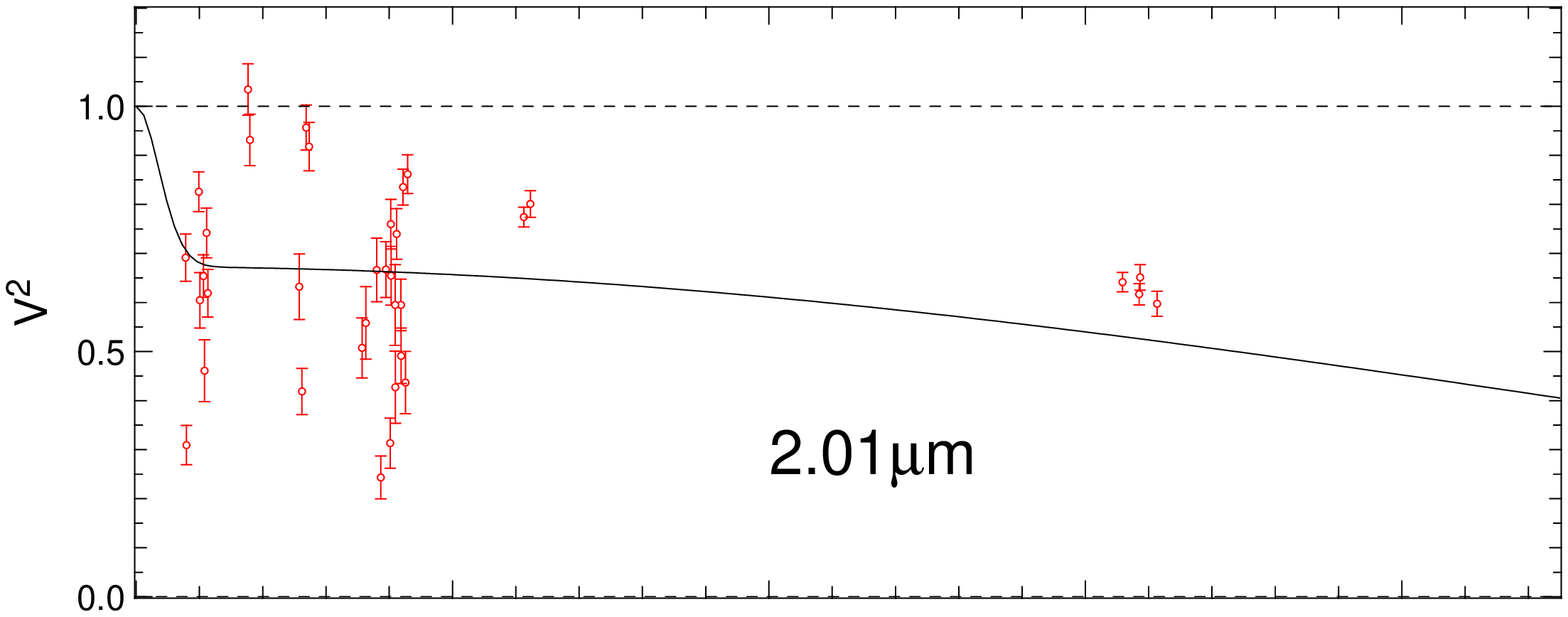}
  \includegraphics[height=0.4\textwidth,angle=-90]{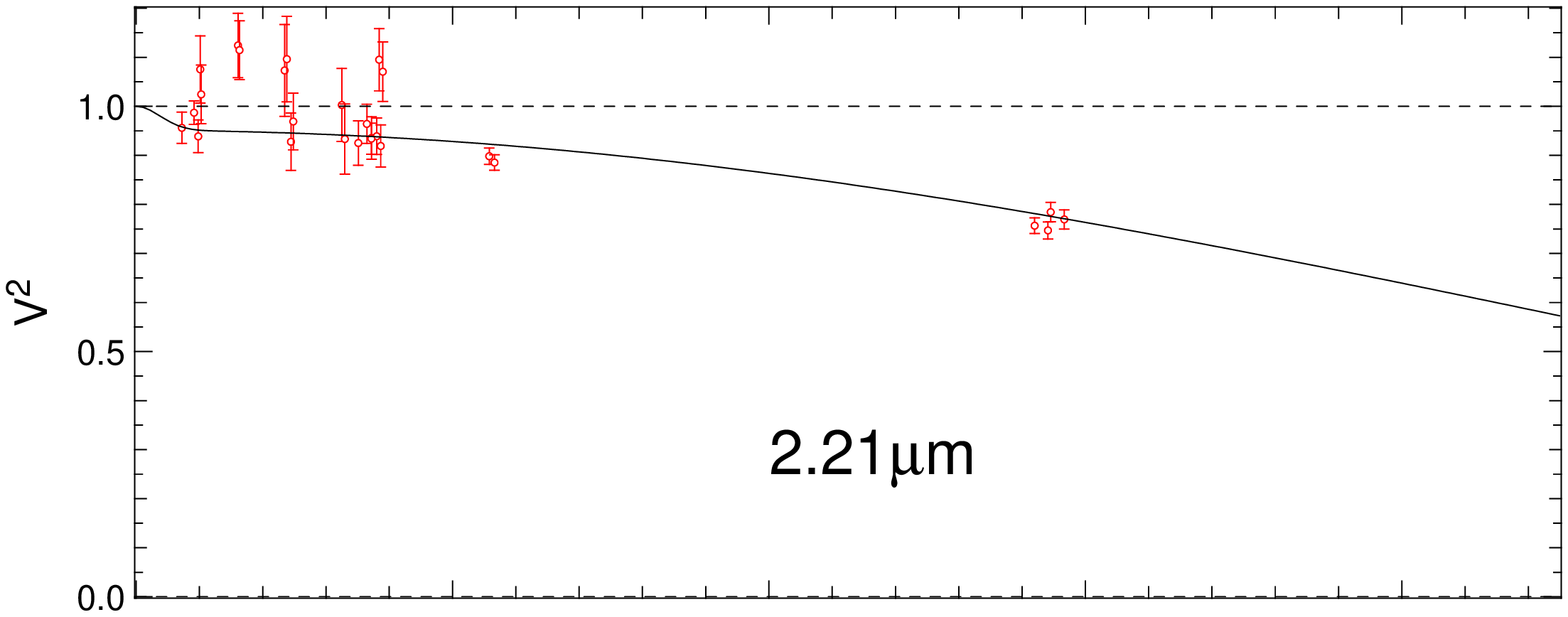}
  \includegraphics[height=0.4\textwidth,angle=-90]{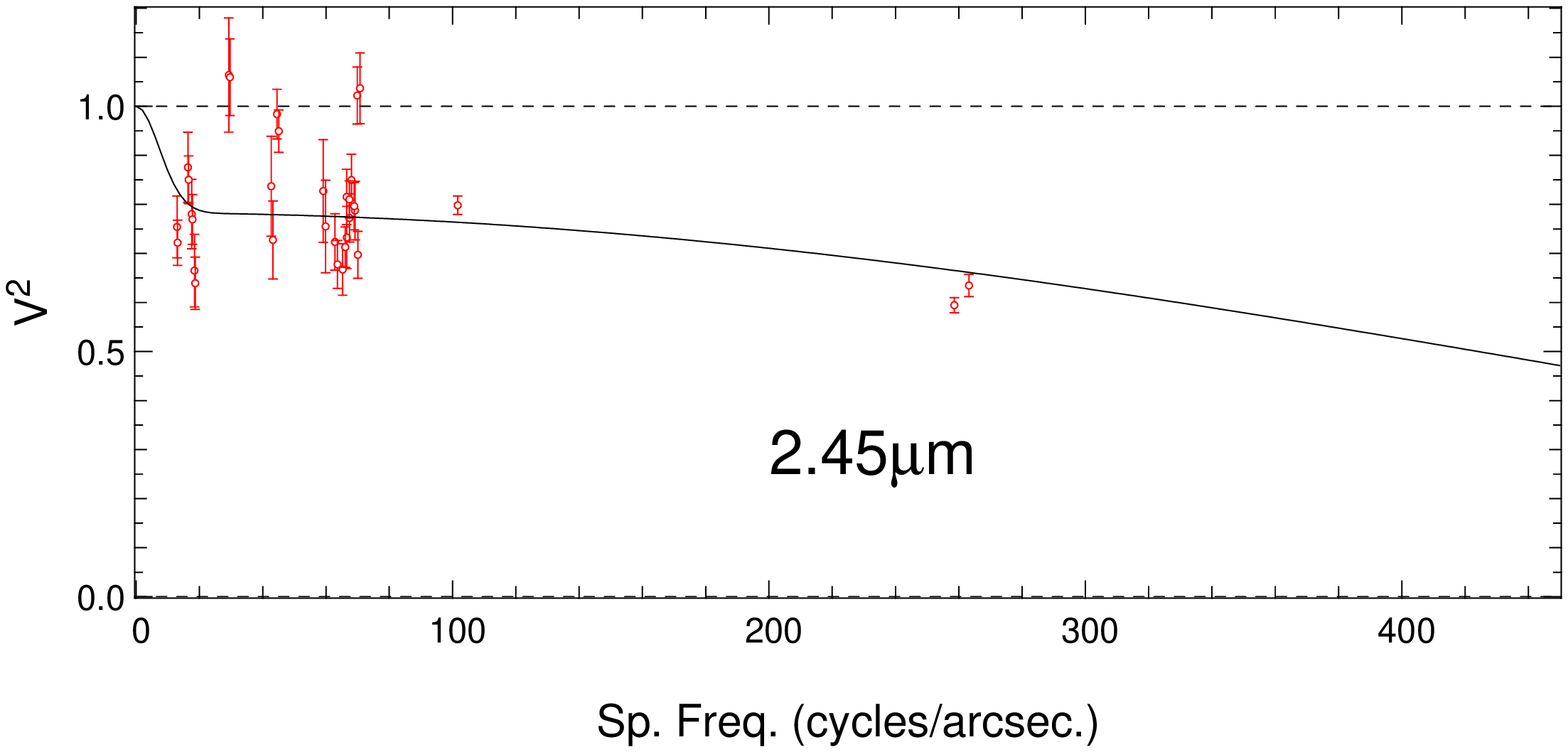}
  \caption{AMBER data shown as a function of spatial frequencies for
    selected wavelengths. The black line shows our best-fit model of a uniform
    disk plus an extended component (50\,mas FWHM here).}
  \label{fig:AMBER_Vis}
\end{figure}

\begin{figure}[htbp]
  \begin{center}
    \centering
    \includegraphics[width=0.45\textwidth]{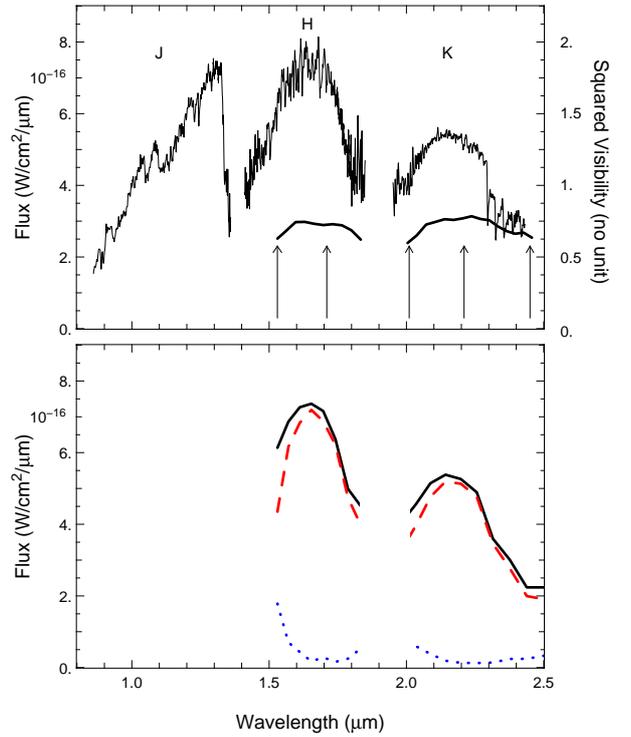}
    \caption[Abu_spectrum]{{\bf Top:} The 0.85--2.5 micron spectrum of
      V838 Mon on 11 December 2013 corrected for extinction using
      E(B-V) = 0.5. For comparison, we show the AMBER longest baseline
      visibility (thick black lines, right axis), and indicate with
      arrows the wavelengths of Fig.~\ref{fig:AMBER_Vis}. {\bf Bottom:
      } Spectra extracted from the AMBER data according to a
      downgraded resolution version of the above spectrum (black
      line). Red dashed line: the star spectrum. Blue dotted line:
      extended emission spectrum.}
    \label{Abu_sp}
  \end{center}
\end{figure}

Visibilities are a proxy to the object's size and shape. The AMBER
visibilities are presented in Fig.~\ref{fig:AMBER_Vis} at selected
wavelengths for all baselines, and for the longest baseline in
Fig.~\ref{Abu_sp} together with the JHK Mt. Abu spectrum. Strong
variations as a function of wavelength are seen, especially at the
edge of the $H$ and $K$ bands where they decrease relative to the
bands centers (1.7\,\micron\ and 2.2\,\micron). All baseline lengths
show this curved shape.
We are confident this is not a flux bias or instrumental effect,
but comes from the source itself.

Closure phases can be qualitatively interpreted as a proxy to
asymmetries: a non-zero (modulo $\pi$) closure phase is a sign of
asymmetries in the object's shape.
The 2014 closure phases are of good quality and compatible with zero
with a mean value of $-1$\fdg$04\pm1$\fdg$12$. The object is therefore
likely centro-symmetric as seen by AMBER. The spectral-range of the
AMBER data is 1.5--2.5\,\micron.

The {{\sc MIDI}} data were processed using the {\tt MIA+EWS}
software\footnote{http://home.strw.leidenuniv.nl/$\sim$nevec/MIDI/}
\citep{2007NewAR..51..666C}. The data were secured in High\_Sens mode
implying that the photometry is obtained subsequently to fringes. The
quality of the data ranges from high quality (error level $\sim$10\%)
to poor quality for some baselines (error level reaching
$\sim$25\%). The spectrophotometric calibration of the mid-IR flux was
performed with the calibrator \object{HD\,52666} (M2III, IRAS
F12=18.3\,Jy with a 5\% error). The ten {\sc MIDI} spectra from two
telescopes were merged leading to a spectrophotometry with an 8\%
error level, representing the mid-IR flux of the source between
October and February 2011. The spectrum was compared with measurements
at the same epoch from the {{\sc WISE}}\footnote{taken in April 2 and
  October 10, 2010} and {{\sc AKARI}}\footnote{obtained between
    May 2006 and August 2007} satellites obtained with much larger
  apertures. The flux in the {{\sc WISE}} filter ($\lambda=
  11.56$\,\micron\,, $\Delta \lambda = 5.51$\,\micron) is
  31.56$\pm$0.06\,Jy and in the {{\sc AKARI}} filter ($\lambda=
  8.22$\,\micron\,, $\Delta \lambda = 4.10$\,\micron) the flux is
  20.11$\pm$0.39\,Jy. The spectrophotometry and the corresponding
  Gaussian estimates from the {{\sc MIDI}} are shown in
  Fig.\ref{fig:MIDI}.


\section{Analysis}
\label{Analysis}

\begin{figure}[htbp]
  \centering
  \includegraphics[width=0.35\textwidth]{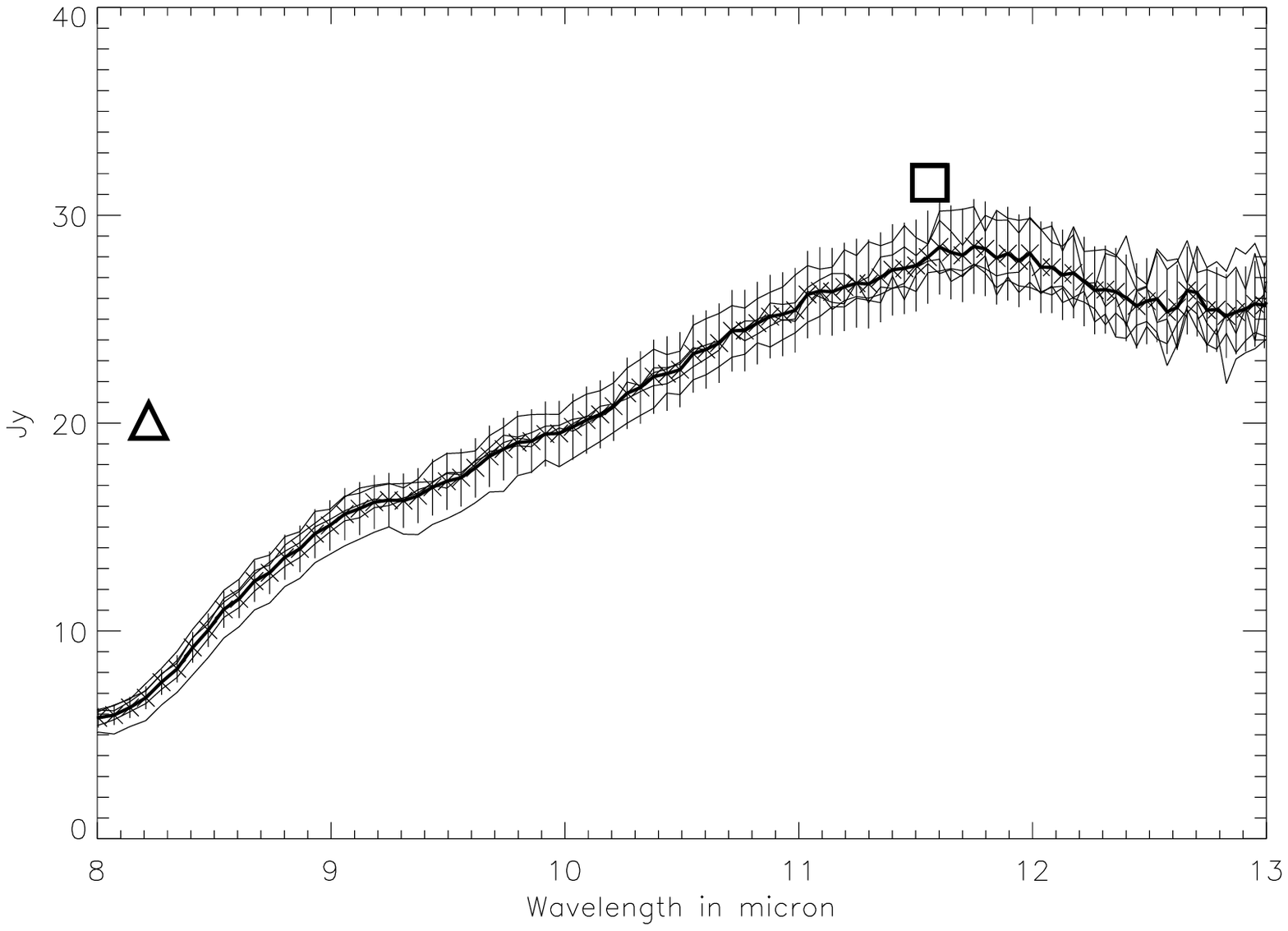}
   \includegraphics[width=0.4\textwidth,angle=0]{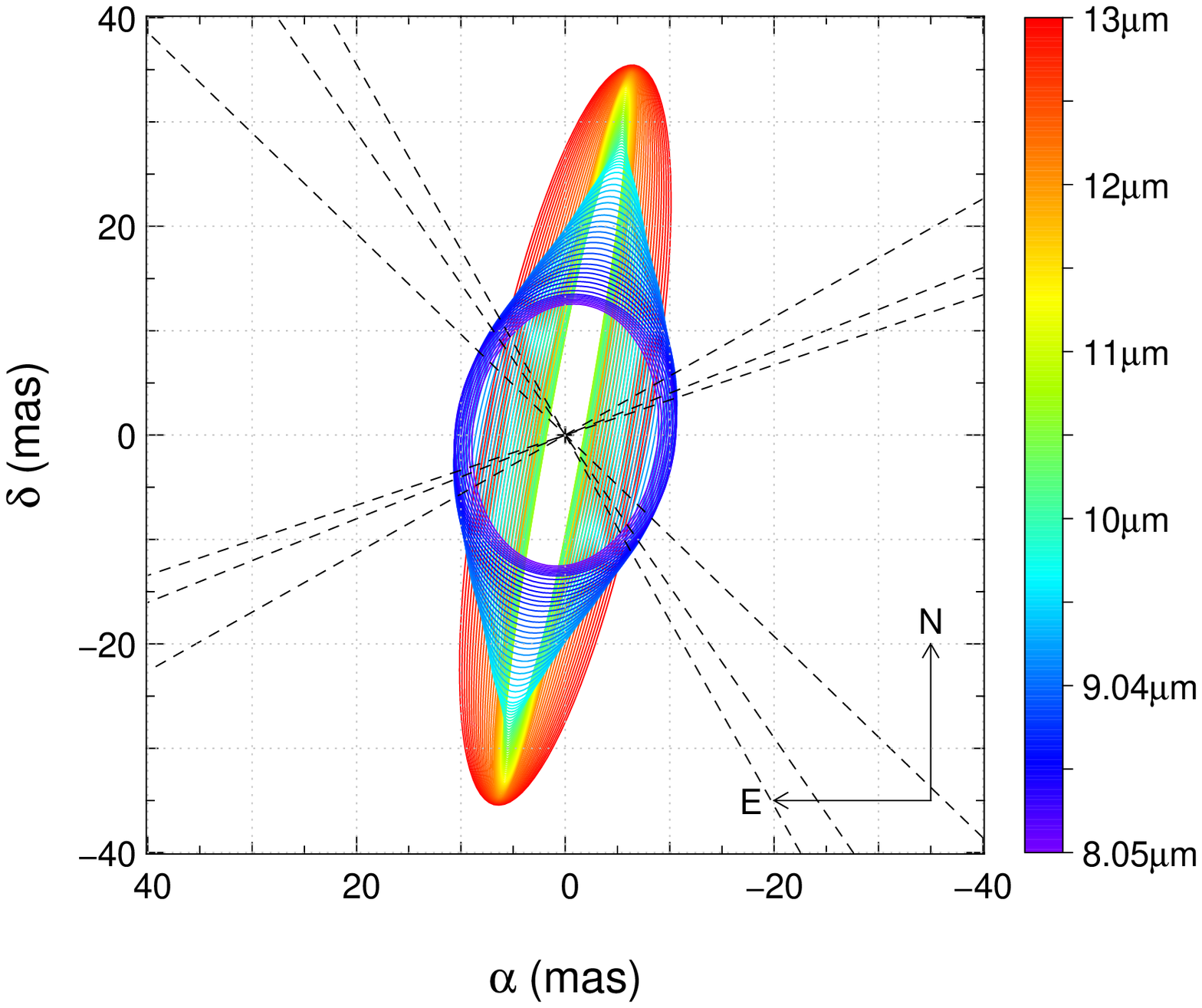}
   \caption{{\bf Top:} {{\sc MIDI}} spectrum obtained in 2011. We
     added for comparison {{\sc AKARI}} (triangle) and {{\sc WISE}}
     (square) fluxes. {\bf Bottom:} 2D Gaussian fit to the whole {{\sc
         MIDI}} dataset. The dashed lines represent the direction of
     the observation baselines. \label{fig:MIDI}}
\end{figure}

The {{\sc MIDI}} dispersed visibilities were translated to the
simplest possible geometrical ad-hoc model -- a Gaussian brightness
distribution, as described in \citet{2004A&A...423..537L}. The
resulting fits shown in Fig.\,\ref{fig:MIDI} depict the general
appearance of the mid-infrared structure. The semi-major and
semi-minor axes of the drawn ellipses represent the half-width at half
maximum (HWHM) of the Gaussian. The extent of that structure increases
from a FWHM of 25\,mas at 8\,\micron\ to 70\,mas at 13\,\micron, with
a high flattening ratio. Despite the heterogeneity of the quality of
the {{\sc MIDI}} visibilities, the flattening of that structure is
certain, but we can only provide a loose range for flattening ratio
values between 1.1 and 2.5 near 9\,\micron, and between 1.3 and 60
near 12\,\micron. The major axis is oriented at a P.A. close to
$-10$\deg\ with an uncertainty of more than $\pm 30$\deg. The
flattening increases with the wavelength, from a roughly round
structure at 8\,\micron\ to a very flattened structure at
13\,\micron. This may come from the contribution of the star, stronger
at 8\,\micron. The flux from {{\sc WISE}} agrees fairly well with the
{{\sc MIDI}} flux. {{\sc AKARI}} data is offset in time from the
  MIDI data by 5 to 6 years. Given that the source's SED has been
  evolving in the IR \citep{2008ApJ...683L.171W} it is not unexpected
  or surprising that there is some difference in the value of the 9
  micron flux measured by AKARI and MIDI.  This implies that most of
the mid-IR flux in 2011 originates from the vicinity of the central
star ($\leq 0.4$'').

The 2013 short-baseline AMBER visibilities are all very close to
unity in the band centers at 1.7\,\micron\ and 2.2\,\micron\ and are
consistent with a completely unresolved object. The long-baseline,
2014 visibilities are close to $V^2=0.8$ at the same wavelengths.

The visibilities as a function of wavelength of the $H$ and $K$
  bands decrease relative to the bands centers (1.7\,\micron\ and
  2.2\,\micron), as shown in Fig.~\ref{Abu_sp}. This is indicative of
  an object whose shape changes as a function of wavelength. At the
  same time, the visibilities at the edge of the bands (1.5\,\micron,
  2.0\,\micron\ and 2.4\,\micron) show a ``plateau-like'' shape as a
  function of spatial frequencies (see Fig.~\ref{fig:AMBER_Vis}),
  indicative of an additional resolved component. We therefore
  analysed first the $K$ band center using a single-component model,
  and then added an additional component to analyze the full AMBER
  dataset.

We used the \texttt{fitOmatic} software \citep{2009A&A...507..317M} to
derive an angular diameter assuming a uniform disk model around
2.2\micron. If we include all the 2013 (short baselines) and 2014
(long baselines) observations, we find a diameter of
$1.17\pm0.37$\,mas. If we keep only the 2014 data, we obtain a
diameter of $1.15\pm0.20$\,mas.
We adopt the latter diameter of $1.15\pm0.20$\,mas as the 2013 data
(short baselines) do not resolve the central star. This measured
diameter is significantly smaller than the one published by
\citet{2005ApJ...622L.137L}.

The AMBER closure phases are all equal to zero within the error
bars. We can therefore constrain the flux of an hypothetical companion
star in addition to the central merger. To achieve that, we added a
point-source to our model and used a simulated annealing algorithm to
constrain its parameters using the observed visibilities and closure
phases and starting with random initial parameters. Our conclusion is
that its flux was always smaller than 3\% of the total flux,
i.e. there would be at minimum 4 magnitudes difference between primary
and secondary companion. The AMBER observations do not exclude the
presence of the companion, but put important constraints on its NIR
contribution to the SED.

The decreasing visibilities at the edges of the $H$ and $K$ bands
indicate the presence of extended emission with a varying spectrum (as
shown in lower panel in Fig.~\ref{Abu_sp}, blue dotted line). We
cannot constrain the exact size of this contribution due to the
plateau-like shape of visibilities as a function of spatial
frequencies, but we can provide a lower limit of its size of
$\theta\geq20$\,mas. We tentatively associate this emission to the
MIDI elongated dusty structure. The extended component would
correspond to the $H$ and $K$ bands contribution of the MIDI elongated
structure. It is noteworthy that adding this additional source of
emission does not significantly change the estimated diameter for the
central star.

To complement these measurements, we used a $JHK$ spectrum of the star
from Mt Abu (India), obtained on 11 December 2013, to derive a
low-resolution spectrum of the central star and the extended emission
(Fig.~\ref{Abu_sp}, lower panel). The stellar spectrum is typical of a
cool luminous star \citep[see ][ for a comparison]{Lancon2000,
  2002A&A...395..161B} with prominent first overtone CO bands at
2.29\,\micron\ and beyond. Similar, albeit weak, CO second overtone
features are also seen in the $H$ band. The deep down-turns at the
band edges are due to the presence of water in its atmosphere
\citep{2005ApJ...627L.141B}. Other molecular features of VO and TiO
are also seen, including some rather uncommon bands of AlO. The
spectrum of the extended structure shows emission bands at the $H$ and
$K$ band edges, where the water absorption occurs in the total
spectrum.

\section{Discussion}
\label{Discussion}

The P.A. of the major axis, as inferred from the dusty flattened
structure discovered by {{\sc MIDI}} is $\approx-10$\deg.
The {\sc MIDI} size increase is expected for a non-truncated dusty
disk. At the adopted distance, the dust is distributed between 150 and
400\,AU from \object{V838\,Mon}.
Such findings are also in agreement with the earliest accurate
spectropolarimetric measurements reported by
\citet{2003ApJ...598L..43W, 2003ApJ...588..486W}. They reported an
intrinsic polarisation with a P.A. of $127$\fdg$0
\pm0$\fdg$5$\ interpreted as scattering by a disk with a major axis at
a P.A. of $37$\deg.

The 2004 {{\sc PTI}} interferometric data had been interpreted as
either a uniform disk or an elliptical model, both consistently
explaining the data well. We note that the P.A.=$15^{+3}_{-27}$\deg
derived from PTI is roughly aligned with our longest AMBER baselines,
i.e. we would be sensitive to the major axis size in case the correct
\citet{2005ApJ...622L.137L} model had been that of an elongated
structure. Our {{\sc AMBER}} data suggest a different picture. The
supergiant has decreased in angular size by $\approx 40$\% in $\approx
10$ years since the measurements of \citet{2005ApJ...622L.137L}. The
linear radius is now estimated to $750 \pm 200$\,\rsun\ ($3.5 \pm
1.0$\,AU). Such a radius is comparable to the famous M2 Iab supergiant
Betelgeuse \citep[$885\pm 90$\,\rsun]{2009A&A...508..923H}. Such an
evolution has already been suggested by \citep{2007A&A...467..269G}. %
This diameter decrease means the star's photosphere has shrinked
during the intervening period at an approximate rate of $1$\kms.

The good quality of the closure phases provides crucial constraints
and excludes any secondary star brighter than a few percent of the
total flux in the 6--600\,AU separation range.
Yet, we caution that a B3V star whose $K$ band magnitude is fainter
than 15 would be totally undetectable by the interferometer.

The {{\sc AMBER}} interferometric observations obtained 10\,yr after
{{\sc PTI}} suggest a contraction of the central star by about 40\%,
which is in line with the spectral change of \object{V838\,Mon}
toward a 'normal' M-type supergiant. No clear sign of deformation of
the photosphere is detected.

The extended emission found in the AMBER data is enhanced relative to
the central star around the CO overtone absorption features and the
H$_2$O absorption at the band edges. This means there is water
emission around the star at an angular scale larger than 20\,mas, and
smaller than 250\,mas (field of view of one telescope at
2.2\,\micron). We tentatively associate this emission to the MIDI
elongated dusty structure.

Several hypotheses could be proposed to explain the detection of the
flattened structure at large distance, as observed by {{\sc MIDI}} and
supported by {\sc AMBER}.

The most likely hypothesis is that the flattened structure is simply
the relic of the large dust formation event as a consequence of the
merging of the two stars \citep{2008ApJ...683L.171W,
  2003ApJ...598L..43W, 2013MNRAS.431L..33N}. The ejecta velocity was
low at outburst time -- less than 200\kms as derived from P-Cygni
profiles by \citet{2004ApJ...607..460L} from several lines. In the 10
years since outburst, ejected material would not have travelled beyond
400\,AU. This scenario is compatible with our finding.

The interferometric observations may also be discussed in the context
of the B3V companion discovered around \object{V838\,Mon}
\citep{2002IAUC.7982....1D, 2002IAUC.7992....2W,
  2002IAUC.8005....2M,2005A&A...434.1107M} and observed in the
post-outburst spectra. The presence of the companion would explain the
X-ray activity discovered in the vicinity of \object{V838\,Mon}
\citep{2010ApJ...717..795A}. It is important to note that the distance
of this companion from the central star was investigated by
\citet{2007A&A...474..585M}, and they estimated the minimum separation
to about 28\,AU. On the other hand, \citet{2009A&A...503..899T}
derived a separation of $\approx250$\,AU. This is well within the
separation range detectable by AMBER. The {{\sc MIDI}} measurements
suggest a large separation of about 100\,AU. In 2009, the B3V hot
signature was no longer observed \citep{2009ATel.2211....1K}. We would
propose then that the companion is building a circumbinary disk via
the Lagrangian point L2, as proposed in some observational and
theoretical studies on other sources \citep{1995A&A...293..363P,
  2010A&A...512A..73M, 2011A&A...526A.107M}.

To conclude, we find a flattened dusty structure around V838
Mon. We also find that the central star decreased its diameter by
nearly 40\% in 10 years.

The interferometric picture of \object{V838\,Mon} as of today is the
following: it is now slowly becoming an anonymous red supergiant,
surrounded by a flattened, probably transitory, dusty environment
extending up to several hundreds of AU.

\begin{acknowledgements}
  O. Chesneau had wished to express his deepest gratitude to the
  hospital staff for their professionalism and devotion during the
  writing of this letter. This letter makes use of the CDS, the JMMC
  and NExScI services.
\end{acknowledgements}

\bibliographystyle{aa}
\bibliography{V838Mon}

\appendix

\section{Bad 2012 calibrator: a newly discovered binary star}
\label{sec:badcal}

\begin{figure}[htbp]
  \centering
  \includegraphics[height=0.4\textwidth,angle=-90]{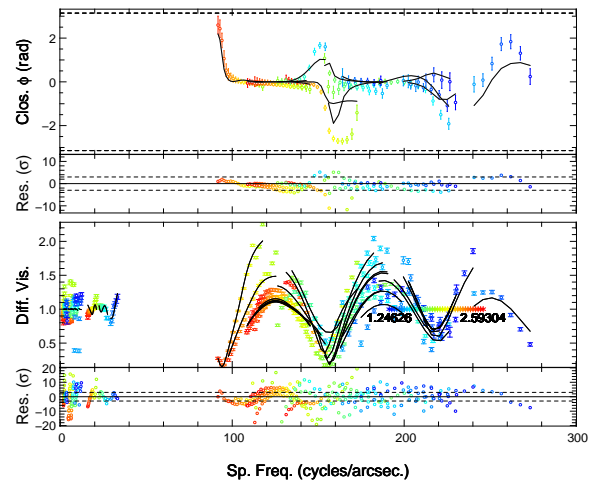}
   \caption{AMBER closure phases and differential visibilities (no
     calibrated $V^2$ available) for HD45299 (color points with error
     bars) together with a best-fit model of a binary star (black
     lines). The spatial frequencies are projected along the binary
     direction ($-67$\deg) to show the binary
     modulation. \label{fig:HD45299}}
\end{figure}

The selected calibrator for the 2012 observations, \object{HD\,45299},
is a previously-unseen visual binary star (see Fig.~\ref{fig:HD45299})
that we detect at 16.0\,mas separation, position angle (P.A.)
$-67$\deg and 0.6/0.4 relative fluxes.

\end{document}